\documentclass[twocolumn,showpacs,preprintnumbers,amsmath,amssymb]{revtex4}

\usepackage{graphicx}% Include figure files
\usepackage{dcolumn}% Align table columns on decimal point
\usepackage{bm}% bold math

\begin{document}

\title{Organization of ecosystems in the vicinity of a
novel phase transition}

\author{Igor Volkov}

\author{Jayanth R. Banavar}
%\email{banavar@psu.edu}
\affiliation{Department of Physics, The
Pennsylvania State University,104 Davey Laboratory, University
Park, PA 16802, USA}

\author{Amos Maritan}
%\email{maritan@sissa.it}
\affiliation{Dipartimento di Fisica `G. Galilei', Universit\`a di
Padova and INFM, via Marzolo 8, 35131 Padova, Italy and The Abdus Salam
International Center for Theoretical Physics (ICTP), Italy}

\date{\today}

\begin{abstract}
It is shown that an ecosystem in equilibrium is generally
organized in a state which is poised in the vicinity of a novel
phase transition.
\end{abstract}


\maketitle

An ecological community consists of individuals of different
species occupying a confined territory and sharing its
resources\cite{MacArthur1,May1,Levin1,Wilson1}. One may draw
parallels between such a community and a physical system
consisting of particles. In this letter, we show that an ecosystem
can be mapped into an unconventional statistical ensemble and is
quite generally tuned in the vicinity of a phase transition where
biodiversity and the use of resources are optimized.

Consider an ecological community, represented by individuals of
different species occupying a confined territory and sharing its
common resources, whose sources (for example, solar energy and
freshwater supplies) depend on the area of the territory, its
geography, climate and environmental conditions. The amount of
these resources and their availability to the community may change
as a result of human activity and natural cataclysms such as
climate change due to global warming, oil spills, deforestation
due to logging or volcanic eruptions.

Generally, species differ from each other in the amount and type
of resources that they need in order to survive and successfully
breed.  One may ascribe a positive characteristic energy intake
per individual, $\varepsilon_k$, of the $k$-th species {\bf($ 0 <
\varepsilon_0 < \varepsilon_1 < \dots $)}, which ought to depend
on the typical size of the
individual\cite{Schmidt1,Calder1,McMahon1,Peters1}. The
individuals of the $k$-th species play the role of particles in a
physical system housed in the energy level $\varepsilon_k$.

We make the simplifying assumption that there is no direct
interaction between the individuals in the ecosystem. Our analysis
does not take into account predator-prey interactions but rather
focuses on the competition between species for the same kind of
resources.

In ecology, unlike in physical systems where one has a fixed
(average) number of particles and an associated average energy of
the system, one needs to define a new statistical ensemble. One
may define the maximum amount of resources available to the
ecological community to be $E_{max}$. The actual energy used by
the ecosystem, $\langle E\rangle$, can be no more than $E_{max}$
\begin{equation} \label{nq1}
E_{max} = T+\langle E\rangle
\end{equation}
and, strikingly, as shown later, the energy imbalance, $T\geqslant
0$, plays a dual role. First, for a given $E_{max}$, the smaller
the $T$, the larger the amount of energy utilized by the system.
Thus, in order to make best use of the available resources, a
system seeks to minimize this imbalance. Second, $T$ behaves like
a traditional `temperature' in the standard ensembles in
statistical mechanics\cite{Feynman1} and, in an ecosystem,
controls the relative species abundance.

\section{Sketch of the derivation}

Consider the joint probability that the first species has $n_1$
individuals, the second species has $n_2$ individuals and so on:
\begin{equation} \label{s2}
P_{eq}(n_1,n_2,...) \propto \prod_k
P(n_k)\Theta(E_{max}-\sum_{j=1}^S\varepsilon_j n_j),
\end{equation}
where $S$ is the total number of species (we assume that $S\gg1$)
and $\Theta(u)$ is a Heaviside step-function (defined to be zero
for negative argument $u$ and $1$ otherwise) which ensures that
the constraint is not violated. Here, $P(n)$ represents the
probability that a species has $n$ individuals in the absence of
any constraint and is the same for all species. Note that $P(n)$
does not have to be normalizable -- one can have an infinite
population in the absence of the energy constraint.

As in statistical mechanics\cite{Feynman1}, the partition function
(the inverse of the normalization factor), $Q$, is obtained by
summing over all possible microstates (the abundances of the
species in our case):
\begin{equation} \label{s3}
Q=\sum_{\{n_k\}} P_{eq}(n_1,n_2,...).
\end{equation}
On substituting Eq.~(\ref{s2}) into Eq.~(\ref{s3}) and
representing the step-function in the integral form one obtains
\begin{eqnarray} \label{s4}
Q=\sum_{\{n_k\}}\prod_k P(n_k) \int_{\gamma} \frac{dz}{2\pi i
z}e^{z(E_{max}-\sum_{j=1}^S\varepsilon_j n_j)}=\nonumber
\\
\int_{\gamma} \frac{d z}{2\pi i z}e^{z E_{max} - \sum_k
h(z\varepsilon_k)},
\end{eqnarray}
where $e^{-h(\beta)} = \sum_n e^{-n\beta}P(n)$ and the contour
$\gamma$ is parallel to the imaginary axis with all its points
having a fixed real part $z_0$ (i.e. $z \in \gamma \Leftrightarrow
z= z_0 +i y , -\infty < y < +\infty$ ). The integral is
independent of $z_0$ provided $z_0$ is positive\cite{Morse1}.

We evaluate the integral in Eq.~(\ref{s4}) by the saddle point
method\cite{Morse1} by choosing $z_0$ in such a way that the
maximum of the integrand of Eq.~(\ref{s4}) occurs when $y=0$:
\begin{equation} \label{s6}
E_{max} - \frac{1}{z_0} = \sum_k\varepsilon_k
h'(z_0\varepsilon_k),
\end{equation}
where the prime indicates a first derivative with respect to the
argument. Note that the rhs is simply $\sum_k\varepsilon_k\langle
n_k\rangle $ with the average taken with the weight $P(n_k)\exp[
-z_0\varepsilon_k n_k]$.

Comparing Eq.~(\ref{s6}) with Eq.~(\ref{nq1}), one can make the
identification $z_0=1/T$ and therefore
\begin{equation} \label{s7}
Q\propto\sum_{\{n_k\}}\prod_k P(n_k)\exp( -\varepsilon_k n_k/T).
\end{equation}
This confirms the role played by the energy imbalance  as the
temperature of the ecosystem. Indeed, the familiar Boltzmann
factor is obtained independent of the form of  $P(n_k)$. Note that
Eq.~(\ref{s7}) with $P(n)=1$  leads to a system of non interacting
identical and indistinguishable Bosons, whereas $P(n)= 1/n!$
describes distinguishable particles obeying Boltzmann statistics.
It is important to note that terms neglected in the  evaluation of
the integral in Eq.~(\ref{s4}) contribute for finite size systems
but these  vanish in the thermodynamic limit.

In order to derive an expression $P(n)$,  consider the dynamical
rules of birth, death and speciation in an ecosystem. In the simplest
scenario\cite{Hubbell1}, the  birth and death rates per individual
may be taken to be independent of the population of the species,
with the ratio of these rates denoted by $x$. Furthermore, when a
species has zero population, we ascribe a non-zero probability of
creating an individual of the species (speciation). Without loss
of generality, we choose this probability to be equal to the per
capita birth rate.

One may write down a master equation for the
dyna\-mics\cite{vanKampen1,Volkov1,Harte1} and show that the
steady state probability of having $n$ individuals in a given
species, $P(n)$, is given by the distribution:
\begin{equation} \label{s1}
P(n)=P(0)\frac{x^n}{n}, n=1,2,3,...
\end{equation}
When $x$ is less than $1$, this leads to the classic Fisher
log-series distribution\cite{Fisher1} for the average number of
species having a population $n$, $\phi(n) \propto P(n)$.

On substituting Eq.~(\ref{s1}) into Eq.~(\ref{s7}), one obtains
\begin{equation} \label{s71}
Q\propto\sum_{\{n_k\}}\prod_k \frac{[x\exp( -\varepsilon_k
/T)]^{n_k}}{n_k},
\end{equation}
where the term $[x\exp( -\varepsilon_k /T)]^{n_k}/n_k$ is replaced
by $1$ when $n_k = 0$.   Note that this leads to an effective
birth to death rate ratio equal to $x\exp( -\varepsilon_k /T) < 1$
for the $k$-th species. In a non-equilibrium situation, such as an
island with abundant resources and no inhabitants, the ratio of
births to deaths can be bigger than $1$ leading to a build-up of
the population. In steady state, however, the deaths are balanced
by births and speciation (creation of individuals of new species).

It is interesting to consider an ecosystem with an additional
ceiling on the total number of individuals, $N_{max}$, that the
territory can hold. In analogy with physics, one may define a
chemical potential\cite{Feynman1}, $\mu \leqslant 0$, so that its
absolute value is the basic energy cost for introducing an
individual into the ecosystem. Thus the total energy cost for
introducing an additional individual of the $k$-th species into
the ecosystem is equal to $\varepsilon_k - \mu$ -- effectively,
all the energy levels are shifted up by a constant amount equal to
this basic cost.

The chemical potential may also be defined as the negative of the
ratio of the energy imbalance to the population imbalance:
\begin{equation} \label{s11}
\mu=-\frac{E_{max}-\langle E\rangle}{N_{max}-\langle
N\rangle}=-\frac{T}{N_{max}-\langle N\rangle},
\end{equation}
where $\langle N\rangle$ is the average population. It follows
then that
\begin{equation} \label{s10}
N_{max}=-\frac{1}{\ln(\alpha)}+\langle N\rangle,
\end{equation}
where $\alpha=\exp(\mu/T)\leqslant 1$. This equation has the same
structure as Eq.~(\ref{nq1}). Interestingly, the link between the
population imbalance and the chemical potential can also be
established formally starting from Eq.~(\ref{s2}), but with an
additional ceiling on the total number of individuals. The
introduction of the ceiling on the population leads to an
additional suppression of the effective birth to death rate ratio
which now becomes $\alpha x\exp( -\varepsilon_k /T)$.

Following the standard methods in statistical
me\-chanics\cite{Feynman1}, one can straightforwardly deduce the
thermodynamic properties by taking suitable derivatives of
$F\equiv -T\ln Q$, the free energy:
\begin{equation} \label{s8}
F=-T\sum_k \ln[1-\ln(1-\alpha x\exp(-\varepsilon_k/T))].
\end{equation}
The average number of individuals in the $k$-th species, $\langle
n_k\rangle\equiv\frac{\partial F}{\partial\varepsilon_k}$, is
given by
\begin{equation} \label{s801}
\langle n_k\rangle=
\frac{\alpha x e^{-\varepsilon_k/T}}{[1-\alpha x e^{-\varepsilon_k/T}]
[1-\ln(1-\alpha x e^{-\varepsilon_k/T})]}.
\end{equation}
In ecological systems, one would expect, in the simplest scenario,
that there ought to be a co-existence of all species in our model
with an infinite population of each when there are no constraints
whatsoever or equivalently when $E_{max}=N_{max}=\infty$. Noting
that $\alpha=1$ when $N_{max}=\infty$, this is realized only when
$1-x \exp(-\varepsilon_k/T)=0$ for any $k$, which, in turn, is
valid if and only if $T=\infty$ and $x=1$.  We will restrict our
analysis\cite{note} in what follows to the case of $x=1$.

Following the treatment in physics\cite{Feynman1}, we postulate
that the number of energy levels, or equivalently the number of
species, per unit energy interval (the density of states) is
proportional to the area of the ecosystem  and additionally scales
as $\varepsilon^d$ with $d>0$ in the limit of small $\varepsilon$.
This is entirely plausible\cite{Schmidt1,Calder1,McMahon1,Peters1}
because one would generally expect the density of states to have
zero weight both below the smallest energy intake and above the
largest intake and a maximum value somewhere in between.

In a continuum formulation, one obtains the following expressions
for the average energy $\langle E\rangle$ and population $\langle
N\rangle$ of the ecosystem:
\begin{equation} \label{nq2}
\langle E\rangle\equiv\sum_{k=1}^S\varepsilon_k\langle
n_k\rangle = T^{d+2}I_1(\alpha)
\end{equation}
and
\begin{equation} \label{nq3}
\langle N\rangle\equiv\sum_{k=1}^S\langle n_k\rangle =
T^{d+1}I_0(\alpha),
\end{equation}
where
\begin{equation} \label{s9}
I_m(\alpha)=\int_0^\infty \frac{\alpha e^{-t}}{[1-\alpha
e^{-t}][1-\ln(1-\alpha e^{-t})]}t^{d+m} dt.
\end{equation}
Note that, except for the second factor in the denominator, which
is subdominant, these integrals ($I_0(\alpha)$ and $I_1(\alpha)$)
are identical to those of a Bose system. The key point is that (in
a Bose system and here) they are both $0$ when $\alpha =0$ and
monotonically increase to their separate finite maximum values at
$\alpha = 1$.

\section{Results and conclusions}

For a given $E_{max}= E_M$, Eq.~(\ref{nq1}) can only be satisfied
over a finite range of temperatures.  The temperature cannot
exceed $T_{max} = E_M$, because $\langle E\rangle$ cannot be
negative.  At this temperature, $\langle E\rangle=\alpha = \langle
N\rangle=N_{max} =0$ and the territory is bereft of life. Also,
the lowest attainable temperature, $T_{min}$, corresponds to
$N_{max}=\infty$ and satisfies the equation
\begin{equation} \label{rrr}
E_M=T_{min}+T_{min}^{d+2}I_1(1)
\end{equation}
(recall that $I_1$ is largest when $\alpha=1$).

The increase of $\langle E\rangle$ on decreasing $T$ is
counter-intuitive from a conventional physics point of view. The
simple reason for this in an ecosystem is that a decreasing $T$
corresponds to a decreasing energy imbalance (first term on rhs of
Eq.~(\ref{nq1}) thereby leading to a corresponding increase in the
second term, which is $\langle E\rangle$. This increased energy
utilization, in turn, leads to an increase in the population of
the community.

\begin{figure}
\includegraphics[width = 0.45\textwidth]{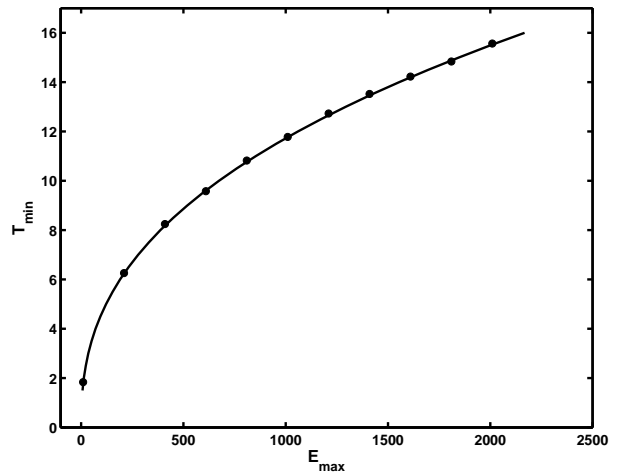}
\caption{\label{fig:f2} Comparison of the results of computer
simulations of an ecosystem with theory. We consider a system with
$100,000$ energy levels with $\varepsilon_k=k^{2/3}$,
$k=1..100,000$, corresponding to $d=1/2$. We work with a constant
$E_{max}$  (the figure shows the results for several values of
$E_{max}$) and consider a dynamical process of birth and death. We
have verified that the equilibrium distribution is independent of
the initial condition. At any given time step, we make a list of
all the individuals and the empty energy levels. One of the
entries from the list is randomly picked for possible action with
a probability proportional to the total number of entries in the
list. Were an individual to be picked, it is killed with $50$\%
probability or reproduced (an additional individual of the same
species is created) with $50$\% probability provided the total
energy of the system does not exceed $E_{max}$. When an empty
energy level is picked, speciation occurs with $50$\% probability
and a new individual of that species is created provided again the
energy of the system does not exceed $E_{max}$. With $50$\%
probability, no action is taken. This procedure is iterated until
equilibrium is reached. The effective temperature of the ecosystem
is defined  as the imbalance between $E_{max}$ and the average
energy of the system (Eq.~(\ref{nq1})). The figure shows a plot of
the effective temperature of the ecosystem deduced from the
simulations. The circles denote the data averaged over a run of
$10^9$ time steps with the last $500$ million used to compute the
average temperature while the solid line is the theoretical
prediction.}
\end{figure}

\begin{figure}
\includegraphics[width = 0.45\textwidth]{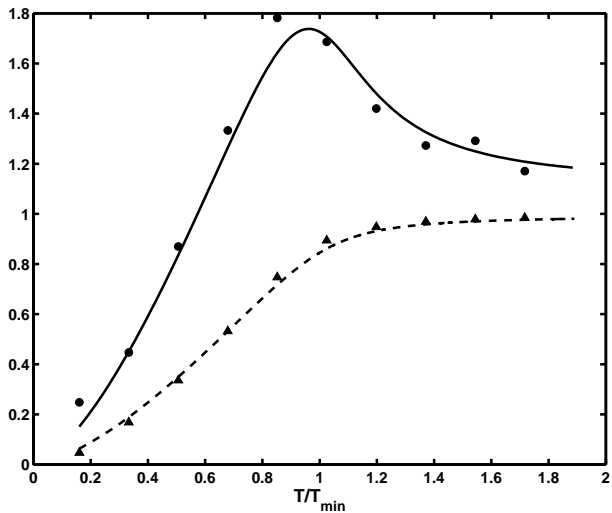}
\caption{\label{fig:f3} Phase transition in an ecosystem with
$N_{max}=N_M= 65$ and $d=1/2$. The dashed and solid curves are
plots of theoretical predictions of $\frac{\langle N\rangle}{N_M}$
and $\frac{\partial \langle E\rangle}{\partial
T}\frac{T_{min}}{E_M}$ respectively versus scaled temperature
$T/T_{min}$, where $T_{min}=11.6$ and $E_M=1000$. The data points
denote the results of simulations.  $\partial \langle
E\rangle/\partial T$ is a quantity analogous to the specific heat
of a physical system and has the familiar $\lambda$ shape
associated with the superfluid transition in liquid
helium\cite{Feynman1}. It was obtained in the simulations as the
derivative of the interpolated values of $\langle E\rangle$. The
continuous phase transition is signaled by the peak in $\partial
\langle E\rangle/\partial T$ (and the corresponding drop in
$\langle N\rangle$) on lowering the temperature and occurs in the
vicinity of the temperature $T_{min}$ (the transition temperature
moves closer to $T_{min}$ as the system size increases).}
\end{figure}

We have carried out detailed computer simulations
(Figs.~\ref{fig:f2} and \ref{fig:f3}) of systems with constraints
on the total energy and the total population and find excellent
accord with theory. Fig.~\ref{fig:f2} shows the results of
simulations corresponding to the case $\alpha=1$ with a constraint
just on the total energy of the ecosystem.

%We have carried out detailed computer simulations to confirm our
%analytic results. Fig.~\ref{fig:f2} shows the excellent agreement
%between the theoretically deduced value of $T_{min}$ and that
%obtained in our simulations confirming that the ecosystem
%organizes itself at this temperature.

A strong hint that there is a link between the behaviors of the
ecosystem at $T_{min}$ and the physical system  of Bosons at the
BEC transition is obtained by noting that both these situations
are characterized by $\alpha = 1$ or equivalently a zero basic
cost ($\mu = 0$) for the introduction of an individual or a
particle into the system.

Let us set the value of $N_{max}$ equal to $N_M$, the average
population in a system with $E_{max}=E_M$ and $N_{max}=\infty$,
and consider the effect of varying the temperature. As in Bose
condensation, one can identify (for $d>0$) a critical temperature,
$T_c$, as the lowest temperature above which the first term on the
rhs of Eq.~(\ref{s10}) can be neglected\cite{Feynman1}. At $T_c$,
$\langle N\rangle\approx N_{max}=N_M$ and $\alpha\approx1$.
Recall, however, that with only the energy ceiling, at $T=T_{min}$
the ecosystem was characterized by $\alpha=1$ and $\langle
N\rangle=N_M$. This confirms that $T_c=T_{min}$. The macroscopic
depletion of the community population, when $T < T_{min}$ is
entirely akin to the macroscopic occupation of the ground state in
BEC (Fig.~\ref{fig:f3}).

When $T$ is larger than $T_{min}$, the energy resources are
sufficient to maintain the maximum allowed population and $\langle
N\rangle = N_M$. Physically, the state at $T_{min}$ corresponds to
a maximally efficient use of the energy resources so that any
decrease in $T$ below $T_{min}$ inevitably leads to a decrease of
the population and the biodiversity in the community. The
existence of this novel transition is quite general and is
independent of specific counting rules such as the ones used in
the classic examples of distinguishable and indistinguishable
particles\cite{Feynman1}.

Our results suggest that ecosystems in equilibrium are organized
in the vicinity of a novel critical point. Critical
points\cite{Wilson2} are generally characterized by large
fluctuations and a slowing down of dynamics in their vicinity. It
is an intriguing possibility that the ubiquity of power laws in
ecology\cite{Schmidt1,Calder1,McMahon1,Peters1} may have some
relationship to our findings here.

\begin{acknowledgments}
We are indebted to Joel Lebowitz for valuable discussions and to
an anonymous referee for useful suggestions. This work was
supported by grants from COFIN MURST 2001, NASA and NSF.
\end{acknowledgments}

\bibliography{bibliography}% Produces the bibliography via BibTeX.

\begin{thebibliography}{17}
\bibitem[{not()}]{note}
\bibinfo{note}{{Our model also shows interesting behavior for $x$ values
  different from $1$. When $x <1$, the birth attempts are fewer than the death
  attempts. In this case, the system is sparsely occupied. On the other hand,
  when $x>1$, one finds, generally, for sufficiently large $E_{max}$ (and
  $N_{max}=\infty$) that the occupancy of the excited levels is small and
  independent of $E_{max}$ with the population of the ground state increasing
  proportional to $E_{max}$. This arises from Eq.~(\ref{s801}): $\langle
  n_1\rangle\rightarrow\infty$ and $\langle n_2\rangle,\langle n_3\rangle,...$
  are finite as $T\rightarrow \varepsilon_1/\ln(x)$ from below.}}
\end{thebibliography}

\end{document}